\documentclass[letterpaper,twocolumn,preprintnumbers,amsmath,amssymb,showpacs]{revtex4}

\usepackage{notation-hein}
\usepackage{qc-people}
\usepackage{fixpapersize}

\newcommand{\lhv}{\ensuremath{\textup{lhv}}}
\newcommand{\adm}[1]{\ensuremath{\operatorname{adm}({#1})}}
\newcommand{\adv}[2]{\ensuremath{\operatorname{adv}_{#1}({#2})}}
\renewcommand{\vec}{}
\iffalse
\newcommand{\fix}[1]{#1}
\else
\newcommand{\fix}[1]{}
\fi

\newtheorem{theorem}{Theorem}
\newtheorem{corollary}{Corollary}
\newtheorem{lemma}[theorem]{Lemma}
\newtheorem{definition}{Definition}
\newenvironment{proof}[1][Proof]{%
  \begin{trivlist}{}{\setlength{\topsep}{0cm}\setlength{\partopsep}{0cm}}
  \item \textbf{#1.\@}\hspace*{1ex}\ignorespaces}%
  {\phantom{.}~\hfill$\Box$\end{trivlist}}

\begin{document}

\title{Multipartite Nonlocal Quantum Correlations Resistant to Imperfections}
\author{Harry Buhrman}
%%\email{buhrman@cwi.nl}
\thanks{Supported in part by the EU fifth framework projects QAIP,
IST-1999-11234, and RESQ, IST-2001-37559.}
\affiliation{CWI and University of Amsterdam,
P.O. Box 94079,
1090 GB Amsterdam, The Netherlands
\addtocounter{footnote}{1}
% above line added to avoid use of \dagger
}
\author{Peter H{\o}yer}
%%\email{hoyer@cpsc.ucalgary.ca}
\thanks{Supported in part by the Alberta Ingenuity Fund and
the Pacific Institute for the Mathe\-ma\-tical Sciences.}
\affiliation{Department of Computer Science, University of Calgary,
             2500 University Drive N.W., Calgary AB, Canada T2N 1N4}
\author{Serge Massar}
%%\email{smassar@ulb.ac.be}
\altaffiliation[Also at]{
Ecole Polytechnique, C.P. 165, 
Universit\'e Libre de Bruxelles, 1050 
Brussels, Belgium}
\thanks{supported in part by the EU fifth framework projects EQUIP,
  IST-1999-11053, and RESQ, IST-2001-37559, by the IUAP program of the
  Belgian Federal Government through grant V-18, and by the Action de
  Recherche Concert\'ee de la Communaut\'ee Fran\c{c}aise de
  Belgique under grant 00/05-251.}
\affiliation{Service de Physique Th\'eorique,
Universit\'e Libre de Bruxelles,
 C.P. 225, Bvd. du Triomphe, 1050
Bruxelles, Belgium} 
\author{Hein R\"ohrig}
\thanks{Supported in part by the EU fifth framework projects QAIP,
IST-1999-11234, and RESQ, IST-2001-37559.}
%%\email{mail@hein.roehrig.name}
\affiliation{Department of Computer Science, University of Calgary,
             2500 University Drive N.W., Calgary AB, Canada T2N 1N4}

\date{October 18, 2004}

\begin{abstract}
  We use techniques for lower bounds on communication to derive
  necessary conditions in terms of detector efficiency or amount of
  super-luminal communication for being able to reproduce with
  classical local hidden-variable theories the quantum correlations
  occurring in EPR-type experiments in the presence of noise.  We
  apply our method to an example involving $n$ parties sharing a
  GHZ-type state on which they carry out measurements and show that
  for local-hidden variable
  theories, the amount of super-luminal classical communication $c$
  and the detector efficiency $\eta$ are constrained by $\eta
  2^{-c/n} = \bigO ( n^{-1/6} )$ even for constant general
  error probability $\epsilon = \bigO(1)$.
\end{abstract}


\maketitle

\section{Introduction}\label{sect:intro}

Forty years ago, Bell~\cite{bell:epr} introduced the notion of 
``quantum nonlocality'': he showed that the correlations between the
outcomes of measurements carried out on entangled quantum systems
cannot be reproduced by a local classical theory (often called a local
hidden variable theory). Since then, extensive
work has been carried out on quantum nonlocality, both on the
experimental and theoretical aspects. On the theory side
research on quantum nonlocality has branched out into many different
and complementary directions.

One important direction of investigation is the search for
qualitatively different types of quantum nonlocality. Of particular
interest is the discovery of the \Greenberger-\Horne-\Zeilinger (GHZ)
paradox~\cite{GHZ,mermin90:unifiedHidden}. 
In this and related examples, correlations are characterized as
nonlocal by the pattern of zero and nonzero joint probabilities.
This property has been called ``pseudo telepathy,'' 
because in every run of the experiment, the parties
appear to agree clandestinely on a subset of admissible outputs. 
It should be
contrasted with other examples where it is the values of these
joint probabilities which implies nonlocality. 

Another important advance was to show that quantum nonlocality
subsists even in the presence of noise as first demonstrated by
\Clauser, \Horne, \Shimony, and \Holt~\cite{CHSH}. This is essential
since every experimental test will necessarily be affected by
imperfections; the best experiments to date have error rates of the
order of a few percent.  Much additional work has been devoted to
understanding the resistance of quantum nonlocality to imperfections.

In experiments involving entangled photons, there is one particular
kind of imperfection that plays a central role, namely the small
efficiency of single-photon detectors.  A single-photon detector will
register the presence of a photon with probability $\eta$, and will
not register the presence of the photon with probability $1-\eta$.
For instance, as one goes from visible to infrared wavelengths, $\eta$
decreases from more than 50\% to 10\%.  Detector inefficiency can be
thought of as a specific type of noise.  This imperfection was first
discussed by \Pearle~\cite{pearle} and remains to
this day one of the major hurdles to overcome in order to carry out a
loophole-free test of quantum nonlocality.  Examples show that there
are quantum correlations that are highly insensitive to detector
inefficiency, but are much more sensitive to other kinds of noise, see
\Massar~\cite{massar:closing}, and therefore this
kind of imperfection should be studied independently of other kinds of
noise. 

Note that the complementary error, namely detectors clicking when they
should not, can also occur. We consider this error as a general noise,
as it cannot be distinguished from other types of noise such as
non-maximally entangled states.

The development of quantum information theory over the past ten
years (see \cite{nielsen&chuang:qc} for a review) has brought a 
breath of fresh air to
the study of quantum nonlocality, and important new questions have
been raised. For instance Bell showed that the quantum
correlations could not be reproduced classically without
``super-luminal'' communication 
between the parties. But \Brassard~\etal~\cite{bct:simulating} and 
Steiner~\cite{steiner:quantifying}
initiated the study of how much super-luminal communication is
required to reproduce the correlations. This question is closely
related to quantum communication complexity, in which one enquires
whether certain distributed communication tasks can be solved using
less quantum communication than is required classically; 
see \cite{buhrman&roehrig:distributedQuantumSurvey} for a survey of
quantum communication complexity.

Remarkably, the amount of classical communication required to
reproduce the quantum correlations and the minimum detector efficiency
required to close the detection loophole are closely related
quantities as demonstrated by \GisinN and \GisinB~\cite{gisingisin},
and \Massar~\cite{massar:closing}.  In many cases, quantum
correlations that require a lot of communication to reproduce
classically cannot be simulated classically without communication,
even when the actual detectors are very inefficient.

Another question that has been raised in the context of quantum
information theory concerns the asymptotic limit when the size of the
entangled system grows. Does the gap between classical and quantum
correlations grow, and if so, at what rate?
\Brassard~\etal~\cite{bct:simulating} showed that in the bipartite
case the amount of communication required to classically reproduce the
quantum correlations can increase exponentially with the number of
entangled bits shared by the parties. And it follows from the results
by \Buhrman~\etal~\cite{bdht:multiparty} that there are quantum
correlations for $n$ parties each holding a two-dimensional subsystem,
so that the amount of communication that must be broadcast in a
classical simulation increases logarithmically with the number of
parties.  Unfortunately these asymptotic results have only been proved
in the total absence of noise.

The only prior asymptotic results in quantum communication complexity
that hold in the presence of noise concern multi-round quantum
communication protocols, such as the appointment-scheduling problem of
\Buhrman~\etal~\cite{BuhrmanCleveWigderson98} or the example due to
\Raz~\cite{raz:qcc}. It appears that these results cannot be mapped to
results concerning quantum nonlocality, whereas communication
complexity problems with a single round of communication and nonlocal
quantum correlations can generally be mapped one onto the other.

The present work lies at the intersection of these different lines of
enquiry. Specifically we concentrate on the generalization of
the GHZ paradox to $n$ parties previously considered by
\Buhrman~\etal~\cite{bdht:multiparty}; their bounds on the
GHZ-inspired multiparty communication problem was only proved in the
absence of noise. We extend it to the noisy case.

These GHZ-type correlations involve $n$ parties. 
We suppose that there is a fixed (\ie independent of $n$)
nonzero probability $\epsilon$ for an error to occur.  
Denote by $c$ the number of bits communicated (via a possibly superluminal
channel) in order to reproduce the correlations. We show that
\[
c = \Omega(n \log n) \enspace .
\]
Denote by $\eta^*$ the maximum detector efficiency for which a local
classical model exists. We show that
\[
\eta^* = \bigO(n^{-1/6}) \enspace .
\]
In fact, the superluminal communication and detection
efficiency can be traded one for the other: we combine the
above two results into the following bound:
\begin{equation}
  \label{REL}
  \eta^* 2^{-c/n} = \bigO \left( n^{-1/6} \right)
  \enspace .
\end{equation}
This bound sheds new light on the
relation between these two quantities, which was previously discussed in
\cite{massar:closing,nonlocalcomb}. 
Our result
constitutes to our knowledge 
the first example in which the degree to which the quantum
correlations are nonlocal increases with the size of the entangled
system in the presence of noise and as such constitute a
significant advance in our understanding of quantum communication
complexity and of quantum nonlocality.

The present work builds upon the earlier results of
\cite{bdht:multiparty} and \cite{nonlocalcomb}. As in these references
we rely heavily on techniques and ideas from the field of
communication complexity. The reader unfamiliar with these notions may
consult the book by \Kushilevitz and
\Nisan~\cite{kushilevitz&nisan:cc} for an introduction to classical
communication complexity.

The remainder of this article is organized as follows. In
Section~\ref{sect:deff} we define precisely the main concepts of
nonlocality used in this paper.  In Section~\ref{sect:comb} we
introduce the combinatorial notion of monochromatic rectangles and
prove a general relation between $\eta$, $c$, and $\epsilon$, which
depends on the maximum size of almost monochromatic rectangles. This
general result is of interest in its own right and could be of use
when studying other instances of quantum nonlocality that exhibit
pseudo telepathy. In Section~\ref{sect:appl} we apply the general bound to
the GHZ paradox; the proof of Eq.~(\ref{REL}) is based on an
addition theorem for cyclic groups proved in Section~\ref{sect:add}.
Finally, we discuss our results and open problems in
Section~\ref{sect:concl}.

\section{Nonlocality Definitions}
\label{sect:deff}

Consider the following situation.  There are $n$ spatially separated
parties; party $i$ receives an input $x_i \in \{ 1, \ldots, k \}$ and
produces an output $a_i \in \{ 1, \ldots, \ell \}$.  With $\vec x = (
x_1, \ldots, x_n )$ and $\vec a = (a_1, \ldots, a_n )$, let $P(\vec a
| \vec x)$ denote the probability of output $\vec a$ given input $\vec
x$.  The inputs are distributed according to the probability
distribution $\mu(x)$.  We formalize this situation as follows.
\begin{definition}\label{def:corrprob}\index{correlation!problem}
  An \emph{$(n, k, \ell)$ correlation problem with input distribution
    $\mu$} is a family of probability distributions $P( \cdot | \vec
  x)$ on the ``outputs'' $\{ 1, \ldots, \ell \}^n$, for each ``input''
  $\vec x \in \{ 1, \ldots, k \}^n$ with $\mu(x) > 0$. We denote the
  support of $\mu$ by $D := \{ x : \mu(x) > 0 \}$.
\end{definition}
Note that in nonlocality experiments the distribution $\mu$ should be
a product distribution, otherwise the parties would have trouble
selecting $x$ according to $\mu$ when the measurements take place in
timelike separated regions. On the other hand the mathematical proofs
given below and in particular the example of Section~\ref{sect:appl},
are based on non-product distributions. The way to get around this is
the following: during the nonlocality experiment the inputs are
distributed according to a product distribution $\mu_0$, for instance
the uniform distribution. Then when analyzing the data one first
throws away part of the data in such a way that, for the data that is
kept, the inputs are distributed according to the desired distribution
$\mu$. This can only make the task harder for the local hidden
variable theory, since it does not know before hand which runs will be
kept and which will be thrown away. From now on we let $\mu$ be an
arbitrary (possibly non-product) distribution.

We are interested in correlation problems obtained from measurements
on multipartite entangled quantum states. We define these as follows.
\begin{definition}\label{def:measscenario}\index{measurement!scenario}
  An \emph{$(n, k, \ell)$ measurement scenario} is a correlation
  problem in which the parties share an entangled state $\ket \psi$;
  each input $x_i$ determines a \mention{positive operator valued
    measure} (POVM) $\hat x_i = \{ \hat x_i^1, \ldots ,\hat x_i^\ell
  \}$ with $\hat x^j_i \geq 0$, $\sum_{j=1}^\ell \hat x^j_i =
  \id_i$.  If the measurement of party $i$ produces outcome $\hat
  x^j_i$, then it outputs $a_i = j$.  The probability
  $P_{\textup{QM}}(\vec a | \vec x)$ to obtain outcome $\vec a$ given
  input $\vec x$ is
  \begin{equation*}
    P_{\textup{QM}}(\vec a | \vec x)
    = \langle \psi | \hat x_1^{a_1} \tensor \cdots \tensor 
    {\hat x}_n^{a_n} | \psi \rangle
    \enspace .
  \end{equation*}
\end{definition}

Our aim is to study what classical resources are required to reproduce
such measurement scenarios.  Let us first consider classical models in
which the parties cannot communicate after they have received the
inputs. Such models are called \define{local}.  The best the parties
can do in this case is to randomly select in advance a deterministic
strategy. This motivates the following definition.
\begin{definition}\label{def:detlhv}
  \index{local hidden variable!model}
  \index{local hidden variable!deterministic model}
  \index{lhv}
  A \emph{deterministic local hidden variable (\lhv{})
    model}\index{local hidden variable!model} is a family of functions
  $\lambda= (\lambda_1,\ldots,\lambda_n)$ from the inputs to the
  outputs: $\lambda_i : \{ 1, \ldots, k \} \rightarrow \{ 1, \ldots,
  \ell \}$.  Each party outputs $a_i = \lambda_i(x_i)$.
  
  A \emph{probabilistic \lhv{} model} (or just \emph{\lhv{} model}) is
  a probability distribution $\nu(\lambda)$ over all deterministic
  \lhv{} models for given $(n, k, \ell)$.
\end{definition}
Thus in probabilistic \lhv{} models the parties first randomly choose
a deterministic \lhv{} model $\lambda$ using the probability
distribution $\nu$. Each party then outputs $a_i = \lambda_i(x_i)$.

We also consider classical models with communication. In such models,
the parties may communicate over a possibly \mention{superluminal}
classical broadcast channel in order to reproduce the quantum
correlations $P_{\textup{QM}}$.  Different communication models exist
depending on whether the parties do not have access to randomness,
possess local randomness only, or share randomness.  These notions are
adapted from the corresponding definitions in communication
complexity.
\begin{definition}\index{local hidden variable!model with communication}
  Consider $n$ parties who each receive an input $x_i \in \{ 1$,
  \ldots, $k \}$, communicate over a classical broadcast channel, and
  each produce an output $a_i \in \{ 1,$ \ldots, $\ell \}$.
  
  A \emph{deterministic classical model with communication} is a
  rooted ``communication protocol'' tree $\mathcal P$; each internal
  node $u$ is labeled with the party $i_u \in \{ 1, \ldots, n \}$
  whose turn it is to broadcast a message; each edge $e$ from $u$ to a
  descendant is labeled with a set $\mathcal X_e \subseteq \{ 1,
  \ldots, k \}$ so that the $\mathcal X_e$ form a partition of $\{ 1,
  \ldots, k \}$; each leaf $v$ is labeled with a \lhv{} model
  $\lambda_v$.  An execution of the protocol on input $x$ starts at
  the root of tree; until a leaf is reached, the execution proceeds
  from node $u$ to the descendant of $u$ that is reached via the edge
  $e$ with $x_{i_u} \in \mathcal X_e$. It is understood that the
  choice of the edge is broadcast to all parties so that all parties
  know at each moment at which node the execution is. When the
  execution has reached the leaf $v$, each party $i$ outputs
  $\lambda_{v,i} (x_i)$ and the execution terminates.  If there are
  $m$ leaves and if the number of children of the nodes on the path
  from the root to the final leaf is $t_1, \ldots, t_m$, the number of
  bits broadcast is $c = \lceil \log t_1 \rceil + \cdots + \lceil \log
  t_{m} \rceil$.
  
  A \emph{classical model with shared randomness} is an arbitrary
  probability distribution $\nu(\mathcal P)$ over deterministic
  classical models. An execution of such a model first
  probabilistically selects a deterministic model and then evaluates
  the deterministic model.
  
  In a \emph{classical model with local randomness}, the distribution
  $\nu(\mathcal P)$ is constrained to be a product distribution of the
  individual strategies of the parties.
\end{definition}
Of course, a classical model that always uses $0$ bits of
communication is just a \lhv{} model.

\begin{definition}
  For a correlation problem $P$ with input distribution $\mu$, we
  denote by $ D (P)$, $R(P)$, and $R^{\textup{pub}}(P)$, respectively,
  the minimum number of bits that must be broadcast in order to
  perfectly reproduce the correlations $P$ when the parties are
  deterministic, have local randomness only, or have shared
  randomness.
  
  Where the choice of the correlation problem $P$\/ is clear from the
  context, we drop it and write $ D$, $R$, and $R^{\textup{pub}}$.
\end{definition}
Clearly, $D(P) \geq R(P) \geq R^{\textup{pub}} (P)$.  Since the results
of quantum measurements are inherently random, it is in 
in general
impossible to
reproduce the quantum correlations using deterministic \lhv{} models
or using deterministic models with communication.  Thus $D(P)$ is
meaningless when trying to simulate quantum measurement scenarios.
However, deterministic models are a very useful tool for studying the
probabilistic models because properties of \emph{all} deterministic
models necessarily also hold for \emph{all} probabilistic models,
since
the probabilistic models are just probabilistic mixtures of
deterministic models. Note also
that \Massar~\etal~\cite{mbcc:siment} showed that $R(P)$ can be
infinite when $P$ arises from a quantum measurement scenario.
In general, classical models cannot reproduce the quantum correlations
$P_{\textup{QM}}$ unless communication is possible, the
detector efficiency $\eta$ is sufficiently small, or 
the error probability is sufficiently large.

Let us consider now the situation where the detectors
are inefficient.
In this case we enlarge 
the space of outputs
to $a_i \in \{ 1, \ldots, \ell \}\cup \{ \perp \}$, where $a_i =
\perp$ is the event that the $i$\/th detector does not produce an
output (``\mention{click}''). We suppose that each measurement $\hat
x_i$ has probability $\eta$ of giving a result and a probability
$1-\eta$ of not giving a result.  Whether a detector clicks or does
not click is independent of the other detectors. This affects the
probabilities in a more structured way than simply decreasing the
probability that all detectors click simultaneously. This issue has
been discussed by \Massar and \Pironio~\cite{massar-pironio}; for
simplicity we will consider here only the two extreme cases, namely
that all detectors click (which occurs with probability $\eta^n$) or
that at least one detector does not click.  We define detector
efficiency accordingly.
\begin{definition}
  Let $P( \cdot | \vec x)$ be a fixed $(n, k, \ell)$ correlation
  problem with input distribution $\mu$.  Let 
  \begin{equation*}
    C := \{ a : \forall i \; a_i \ne \perp \}
  \end{equation*}
  denote the output vectors where all detectors click. With slight
  abuse of notation, we also use $C$ as the indicator random variable
  of the event $a \in C$.  We define the
  \define[detection!efficiency]{detection efficiency} 
$\eta$ of the
  correlations to be the expectation
  \begin{equation*}
    \eta := \left( \expect_{\mu} \left[ \sum_a P(a|x) C \right]
      \right)^{1/n}
    \enspace .
  \end{equation*}
\end{definition}
Note that here the atomic events are tuples $(x,a)$ of an input and an
output vector with a joint distribution of the form $\Pr[ \text{input
} x \text{ and output } a ] = \mu(x) P(a | x)$.  The expectation above
is over the marginal distribution $\mu$ of the inputs.

We are also interested in the possibility that the \lhv model makes
errors.
\begin{definition}
  Suppose that some classical model produces a probability
  distribution $P(a|x)$, which should approximate the probability
  distribution produced by a measurement scenario $P_{QM} (a|x)$. The
  \define{\tvd} is a measure for how much these two distributions
  differ:
  \begin{align*}
    \epsilon_{\textup{var}} := \expect_\mu \left[ \sum_a \left|
        P_{QM}(a|x) - P(a|x) \right| \frac{ C }{ \eta^n} \right] 
  \end{align*}
\end{definition}
The inclusion of the factor $C /
\eta^n$ takes care of the possible finite efficiency of the detectors,
assumed to be the same for $P_{QM} (a|x)$ and for $P(a|x)$.
  
We will be particularly interested in quantum correlations that
exhibit ``pseudo telepathy'', \ie such that $P_{QM} (a|x)=0$ for some
$a$ and $x$. For such correlations it is convenient to define the
error probability as follows.
\begin{definition}
  Let 
  \begin{equation*}
    F := \left\{ (a,x) : P_{QM} (a|x) = 0 \right\}
  \end{equation*}
  and again we also denote by $F$ the indicator random variable of the
  event $P_{QM} (a|x) = 0$.  The \define{error probability} is
  \begin{equation*}
    \epsilon := \expect_\mu \left[ \sum_a 
      P(a|x) F  \frac{ C }{ \eta^n} \right] \enspace .
  \end{equation*}
\end{definition}
Thus $\epsilon$ is the probability to observe in one run an event that
cannot occur in the quantum mechanical model.  It is immediate to
check that
\begin{equation*}
  \epsilon_{\textup{var}} \geq \epsilon \enspace .
\end{equation*}

For an $(n, k, \ell)$ correlation problem $P( \cdot | \vec x)$ with
input distribution $\mu$, we denote by $\eta^*$ the maximum detector
efficiency of any \lhv{} model that reproduces the quantum
correlations, and by $\eta^*_{\epsilon}$ the maximum detector
efficiency that reproduces the quantum correlations up to error
$\epsilon$.  Similarly, we can define $D_\epsilon$, $R_\epsilon$,
$R^{\textup{pub}}_\epsilon$ the amounts of communication required to
reproduce the correlation problem $P$ in the presence of error. We are
interested in $\eta^*_\epsilon$ and by $R^{\textup{pub}}_\epsilon$.

We can map every communication model with $c$ bits of communication
with shared randomness into a model with inefficient detectors with
efficiency $ \eta^n = 2^{ -c } $: the shared randomness determines the
conversation between the parties. Thus they all agree on the
conversation. Each party $i$ checks whether its input $x_i$ is
compatible with the conversation and, if yes, produces output $a_i$
according to the communication model and otherwise produces no output,
\ie $\perp$.  The total probability that all detectors click is equal
to the probability that $\vec x$ belongs to the conversation.  Since
each input belongs to one and only one conversation, the probability
that all detectors click is equal to one over the number of
conversations.  Note that in this model the probability that a
specific detector, say detector $i$, clicks may depend on the input
$x_i$. However, the probability that all detectors click remains
independent of the input.
\begin{theorem}\label{thm:outputForComm}
  Consider \lhv{} models where the probability that all detectors
  click is independent of the input, but where the probability that
  each detector clicks, say detector $i$, may depend on its input
  $x_i$.  Then there exists a \lhv{} model if the probability $\eta^n$
  that all detectors click is at most $2^{-R^{\textup{pub}}}$. This
  implies that in these models,
  \begin{equation}
    ( \eta^* ) ^n \geq 2^{-R^{\textup{pub}}}\enspace .
    \label{model}
  \end{equation}
\end{theorem}
This result was given in \cite{nonlocalcomb} in the absence
of error, but it also holds when errors are present.

\section{Combinatorial bounds}\label{sect:comb}

We now introduce some definitions and notation, which allow us to state
and then prove our result concerning a general relation between $c$,
$\eta$ and $\epsilon$. We are concerned with pseudo-telepathy type
correlations for which there are some $P(a|x)$ that vanish.  
\begin{definition}
  Let $P( \cdot | \vec x)$ be a fixed $(n, k, \ell)$ correlation problem
  with input distribution $\mu$. 
  We define the sets of inputs that
  admit output $a$ as
  \begin{equation*}
    \adm{a} := \{ x : P(a | x) > 0 
    %% \textup{ and } \mu(x) > 0 
    \}
  \end{equation*}
  for all $a \in C$.  Moreover, for a set $S \subseteq \{ 1, \ldots,
  k \}^n$ of inputs and a specific output $a \in \{ 1, \ldots, \ell
  \}^n$, the \emph{$a$-advantage of $S$} is
  \begin{equation*}
    \adv{a}{S} := \frac{ \mu( S \cap \adm{a} ) }{ \mu(S) }
  \end{equation*}
  for all $a\in C$.
\end{definition}

For sets $A_1$, \ldots, $A_n$, a subset $R$ of the Cartesian product
$A_1 \times \cdots \times A_n$ is called a \define{rectangle} if there
are $R_1 \subseteq A_1$, \ldots, $R_n \subseteq A_n$ such that $R =
R_1 \times \cdots \times R_n$, \ie $R$ is a Cartesian product itself.
The importance of rectangles is that for a deterministic \lhv{} model
$\vec \lambda = (\lambda_1, \ldots, \lambda_n)$, the set $ R_{\vec
  \lambda} (\vec a) := \{ \vec x \mathrel{:}\vec \lambda (\vec x) =
\vec a\} $ of all inputs $\vec x$ leading to output $\vec a$ is a
rectangle: $R_{\vec \lambda}( \vec a) = \lambda_1^{-1} ( a_1) \times
\cdots \times \lambda_n^{-1} ( a_n)$.

\begin{theorem}\label{thm:rect}
  Let $P$ be a fixed $(n, k, \ell)$ correlation problem with input
  distribution $\mu$. If for some $\delta$ (\/$0 \le \delta \le 1$), 
  all
  rectangles $R$ with $\adv{a}{R} \ge \delta$  have $\mu(R)
  \le r$ for every $a\in C$, 
  then for every classical model $\nu(\mathcal P)$ with $c$ bits
  of communication holds
  \begin{equation*}
    \frac{1}{2^c} \eta^n \left( 1 - \epsilon \frac{1}{ 1 - \delta}
    \right) \le \ell^n r .
  \end{equation*}
\end{theorem}

This shows the strong relation between the detection
efficiency and the amount of classical communication required to
reproduce the correlations. Indeed one quantity can be traded for the
other. 
\begin{proof}[Proof of Theorem~\ref{thm:rect}]
  Let $R_{\mathcal P, v, a}$ denote the set of inputs $x$ for which
  the deterministic protocol $\mathcal P$ terminates in leaf $v$ and
  outputs $a$. Every $R_{\mathcal P, v, a}$ is a rectangle. Let $L :=
  \{ ( \mathcal P, v, a ) : \adv{a}{ R_{\mathcal P, v, a} } \ge \delta
  \}$.  Then
  \begin{align*}
    \eta^n ( 1 - \epsilon ) &= \sum_{\mathcal P, x} \nu(\mathcal P)
    \mu(x) C (1-F) 
    \displaybreak[2] \\ 
    & = \sum_{\mathcal P, v, a} \nu(\mathcal P) \mu(
    R_{\mathcal P, v, a} \cap \adm{a} ) 
    \displaybreak[2] \\ 
    & = \sum_{\mathcal P, v, a}
    \nu(\mathcal P) \mu( R_{\mathcal P, v, a} )
    \adv{a}{ R_{\mathcal P, v, a} } 
    \displaybreak[2] \\
    & \le \sum_{(\mathcal P, v, a) \in L} \nu(\mathcal P) r +
    \sum_{(\mathcal P, v,
      a) \notin L} \nu(\mathcal P) \mu( R_{\mathcal P, v, a} ) \delta 
    \displaybreak[2] \\
    & \le 2^c d^n r + \delta \sum_{(\mathcal P, v, a) \notin L}
    \nu(\mathcal P) \mu( R_{\mathcal P, v, a} )
  \end{align*}
  where the $v$ range over the leafs of $\mathcal P$ and the $a$ over
  $\{ 1, \ldots, \ell \}^n$.  Similarly,
  \begin{align*}
    \eta^n \epsilon &= \sum_{\mathcal P, v, a} \nu(\mathcal P) \mu(x)
    C F 
    \displaybreak[2] \\ 
    & = \sum_{\mathcal P,v,a} \nu(\mathcal P) \mu \Bigl( R_{\mathcal
      P, v, a} \cap \bigl( \{ 1, \ldots, k \}^n \setminus \adm{a} \bigr) 
    \Bigr) 
    \displaybreak[2] \\ 
    & = \sum_{\mathcal P,v, a}
    \nu(\mathcal P) \mu( R_{\mathcal P, v, a} )
    \bigl(1 - \adv{a}{ R_{\mathcal P, v, a} } \bigr) 
    \displaybreak[2] \\
    & \ge 0 + \sum_{(\mathcal P, v, a) \notin L} \nu(\mathcal P)
    \mu( R_{\mathcal P, v, a } )
    ( 1 - \delta ) 
    \displaybreak[2] \\
    & = (1 - \delta ) \sum_{(\mathcal P, v, a) \notin L}
    \nu(\mathcal P) \mu( R_{\mathcal P, v, a} )
  \end{align*}
  Hence,
  \begin{equation*}
    \eta^n ( 1 - \epsilon ) \le 2^c \ell^n r + \frac{\delta}{1 - \delta} 
    \eta^n \epsilon
    \enspace ,
  \end{equation*}
  which implies Theorem \ref{thm:rect}.
\end{proof}

\section{Application to the GHZ correlations}\label{sect:appl}

In this measurement scenario each of the $n$ parties has a
two-dimensional quantum system. The overall state of the $n$ qubits is
\begin{eqnarray}
  \ket \psi = \frac{| 0^n\rangle + |1^n\rangle}{\sqrt{2}} 
\end{eqnarray}
where $|i^n\rangle=|i\rangle\tensor \ldots \tensor |i\rangle$ with $n$ terms
in the product. Each party receives as input 
$x_i \in \{ 0, \ldots , k-1\}$. Each party then measures his qubit in
the basis 
\begin{eqnarray}
  \ket { \phi_{\pm} } = \frac{|0\rangle \pm e^{ \pi \I x_i / k} |1\rangle}{
    \sqrt{2}}
\end{eqnarray}
If the qubit is projected onto state $\ket{ \phi_+ }$, then party $i$
outputs $a_i=0$ and if the qubit is projected onto state $\ket{
  \phi_- }$, party $i$ outputs $a_i=1$. 

We call an input $x = (x_1,\ldots,x_n)$ \emph{valid} if it satisfies
\begin{equation}\label{eq:promise}
  \left( \sum_{i=1}^n x_i \right) \text{mod $k$} = 0
\end{equation}
and we let $D \subset \integer_k^n$ denote the set of all valid
inputs.  Let $F : \integer_k^n \rightarrow \{0,1\}$ denote the Boolean
function on the valid inputs defined by
\begin{equation*} 
  F(x) = \frac{1}{{k}}
  \left[\left(\sum_{i=1}^n x_i\right) \text{mod $2  k$}\right].
\end{equation*}
The function $F$ can be viewed as computing the $(1 + \log k )$-th
least significant bit of the sum of the~$x_i$.

It is easy to check that the outputs of the quantum measurement are
correlated as follows: if Eq.~(\ref{eq:promise})
holds, then
\begin{equation}\label{PC}
  \left ( \sum_{i=1}^n a_i \right) \bmod 2 = \frac{1}{k}
  \left[ \left ( \sum_{i=1}^n x_i \right) \bmod 2 k \right ] = F(x) 
  \enspace . 
\end{equation}
Hence, if each party broadcasts its measurement outcome then each
party can locally compute $F(x)$.
\begin{lemma}
  In the model with prior entanglement and classical broadcast
  communication, the \mention{communication complexity} of computing
  $F(x)$ is $\bigO (n)$.
\end{lemma}
Moreover, the above measurement scenario will exactly reproduce the
following $(n, k , 2)$ correlation problem (see
Definition~\ref{def:corrprob}): 
\begin{definition}\label{CCC}
Let $\mu (x)$ be a distribution on the
inputs that gives zero weight to the invalid inputs $x$, which do not
satisfy Eq.~(\ref{eq:promise}), and let
\begin{equation*}
  P(a | x) := 
  \begin{cases}
    \frac{1}{2^{n-1}} & \text{if } F(x) = a_1 + \cdots + a_n \mod 2 \\
    0 & \text{otherwise.}
  \end{cases}
\end{equation*}
for all $a \in \{ 0, 1 \}^n$ and $x \in D$.  
\end{definition}
A simple classical strategy for reproducing these correlations is for
every party to broadcast its input. Hence, with $k = n^{1/6}$, the
communication problem and the correlation problem can be solved
exactly with $\bigO ( n \log n )$ bits of communication.  

Note that 
for $n=3$ and $k=2$ the above correlations constitute 
the GHZ paradox as formulated by
\Mermin~\cite{mermin90:unifiedHidden}.  The case $k=2$ and arbitrary $n$
was studied by \Mermin~\cite{mermin:extreme} and recently revisited by \Brassard~\etal~\cite{brassard03:_multi_party_pseud_telep,brassard&broadbent&tapp:recastingMermin}.  In
\Buhrman~\etal~\cite{bdht:multiparty} and our earlier research
\cite{nonlocalcomb} the case where the number of settings $k$
is a power of two was considered.  In \cite{bdht:multiparty} it was
shown that the amount $c$ of classical communication which the parties
must broadcast in order to reproduce exactly the correlations
from Definition~\ref{CCC} is $c = \bigO(n \log n)$ when $k = \bigO(n)$. And in
\cite{nonlocalcomb} it was shown that the maximum detector
efficiency $\eta^*$ for which a local classical model can reproduce
exactly
the correlations from Definition~\ref{CCC} decreases as $1/n$.  Furthermore the
classical strategy described above shows that in the absence of noise
these results are essentially
optimal.

We will now show that this optimality continues to hold in the
presence of noise and that classical strategy described above
remains close to optimal in the presence of noise. 
Specifically we will show that
\begin{theorem}\label{thm:lhvLower}
  Let $\mu$ be the uniform distribution on valid inputs. Then the
  number $c$ of bits broadcast, the efficiency $\eta$ and the error
  $\epsilon$ of every \lhv{} model $\nu$ are constrained by
  \begin{equation*}
    \frac{1}{2^{c/n}} \eta \left ( 1 - \epsilon \left [ 2 + \bigO
        \left ( \frac{1}{n^{1/6}} \right ) \right ] \right )^{1/n} =
    \bigO \left ( \frac{1}{n^{1/6}} \right ) .
  \end{equation*}
\end{theorem}
which, for fixed $\epsilon$, $n$ large, implies Eq.~(\ref{REL}). In
particular we have 
\begin{corollary}\label{cor:CClower}
  \index{communication complexity}%
  Every bounded-error randomized public coin protocol for $F :
  \integer_k^n \rightarrow \{0,1\}$ with $k \geq n^{1/6}$ requires
  $\Omega(n \log n)$ bits of communication.
\end{corollary}

We now turn to the proof of Theorem~\ref{thm:lhvLower}.  We say a
rectangle $R = A_1 \times \cdots \times A_n \subseteq k^n$
\emph{involves $m$ parties\/} if at least $m$ of the $n$ subsets $A_i$
have size at least~2.  Every rectangle involving at most $m$ parties
can have size at most~$k^m$.

\begin{lemma}[Small rectangles are insignificant]\label{lm:minuscule}
  \fix{\hspace{1mm}}Every rectangle $R$ involving at most $n^{5/6}$ parties satisfies
  $\log |R| \leq n^{5/6} \log k
  %% \in o(n \log n)
  $.
\end{lemma}

We say a rectangle $R$ \emph{has bias at most~$\delta$\/} if
\begin{equation*}
  |F^{-1}(1) \cap D \cap R| \leq (1+\delta) |F^{-1}(0) \cap D \cap R|
\end{equation*}
and
\begin{equation*}
  |F^{-1}(0) \cap D \cap R| \leq (1+\delta) |F^{-1}(1) \cap D \cap R|.
\end{equation*}
Note that for every $a$ we have $\adm{a} \cap D = F^{-1} (a_1 + \cdots +
a_n \mod 2) \cap D$. Therefore, if $\mu$ is a distribution that is
uniform on $D$, then $R$ has bias at most $\delta$ if and only if it
has $a$-advantage at most $(1+\delta)/(2+\delta)$ for every $a$.
The next lemma expresses that every ``large'' rectangle is
almost unbiased. 
\begin{lemma}[Large rectangles are almost unbiased]
  \label{lm:unbiased} 
  \fix{\hspace{2mm}}Every rec\-tangle involving at least $n^{5/6}$ parties has bias at most
  $\bigO(1/n^{1/6})$.
\end{lemma}
The proof of Lemma \ref{lm:unbiased} is based on addition theorems for
cyclic groups and is given in the next section.
\begin{proof}[Proof of Theorem \ref{thm:lhvLower}]
  Lemma~\ref{lm:unbiased} implies that each rectangle involving at
  least $n^{5/6}$ parties can have $a$-advantage at most $1/2 +
  \bigO (1/n^{1/6} )$ for any $a$. Hence, rectangles with
  $a$-advantage greater than $1/2 + \bigO (1/n^{1/6} )$
  must involve less than $n^{5/6}$ parties. By
  Lemma~\ref{lm:minuscule}, such a rectangle $R$ has size less than
  $k^{n^{5/6}}$ and thus
  \begin{equation*}
    \mu(R) = |R|/k^{n-1} \le k^{n^{5/6} - n + 1} = n^{- \frac{1}{6} ( n
      - n^{5/6} - 1 ) } .
  \end{equation*}
  Plugging these values into Theorem~\ref{thm:rect}, we obtain
  \begin{equation*}
    \frac{1}{2^c} \eta^n \left ( 1 - \epsilon \left [ 2 + \bigO \left (
          \frac{1}{n^{1/6}} \right ) \right ] \right ) \le 2^{ -
      \frac{1}{6} n \log n + \bigO( n) } 
  \end{equation*}
\end{proof}

\section{an addition theorem}\label{sect:add}

Let $\integer_T$ denote the additive cyclic group of order~$T$.  Let
$\mu_A(x)$ denote the multiplicity of an element~$x$ in the
multiset~$A$.  For multisets $A$ and $B$ of $\integer_T$, let
$A+B$ denote the multiset $\{a+b\,|\, a \in A, b\in B\}$.
\begin{definition}
  We say a multiset $A$ of $\integer_T$ has bias at most~$\epsilon$
  with respect to a subgroup $H \subgroup \integer_T$ if $\mu_A(a)
  \leq (1+\epsilon)\mu_A(a+h)$ for all $a \in A$ and all $h \in H$.
\end{definition}
\begin{theorem}[Addition Theorem]\label{thm:addition}
  Let $A_1, \ldots, A_r$ be subsets of $\integer_T$, each of size at
  least~$2$, with $r \geq T^3$ and $T = 2^{t}$ a power of~$2$.
  Then the multiset $A_1 + A_2 + \cdots + A_r$ has bias at most
  $\bigO({T^{3/2}}/{r^{1/2}})$ with respect to the subgroup
  $\{0,2^{t-1}\}$.
\end{theorem}
Essentially, this theorem is derived by a sequence of simple
reductions to the following observation: We may generate an almost
uniformly distributed random number between $0$ and $K-1$ by flipping
a fair coin $K^2$ times, and counting the number of heads modulo~$K$.
\begin{lemma}\label{fact:adding}
  For multisets $A$ and $B$ over $\integer_T$, if $A$ has bias at
  most~$\epsilon$ with respect to some subgroup~$H$, then so does
  $A+B$.  In particular, the multiset $A+\{d\}$ has the same bias
  as~$A$.
\end{lemma}
\begin{lemma}\label{lm:coins}
  Let $f: \{0,1\}^s \rightarrow \integer_K$ be defined by 
  \begin{equation*}
  f(a_1, \ldots, a_s) = \left( \sum_{i=1}^s a_i \right) \textup{ mod $K$}
  \enspace .
  \end{equation*}
  If $s
  \geq K^2$, then $|f^{-1}(x)| \leq \big(1 + 4 \frac{K}{\sqrt{s}}
  \big) \, |f^{-1}(y)|$ for all $x,y \in \integer_K$.
\end{lemma}
\begin{proof}
  First suppose $x \leq y$.  Then
  \begin{align*}
    \left| f^{-1}(x) \right|
    &=    \sum_{i} \binom{s}{x+iK} \\
    &=    \sum_{i:y+iK< s/2} \binom{s}{x+iK} 
    + \sum_{i:y+iK\geq s/2} \binom{s}{x+iK} \\
    &\leq \sum_{i:y+iK< s/2} \binom{s}{y+iK} \\
    & \phantom{=}\;
    + \sum_{i:y+iK\geq s/2} \binom{s}{x+iK+K} + \binom{s}{s/2}\\
    &\leq \sum_{i:y+iK< s/2} \binom{s}{y+iK} \\
    &\phantom{=}\;
    + \sum_{i:y+iK\geq s/2} \binom{s}{y+iK} + \binom{s}{s/2}\\
    &= \left| f^{-1}(y) \right| + \binom{s}{s/2}.
  \end{align*}
  Similarly, if $x>y$, then still
  $|f^{-1}(x)| \leq |f^{-1}(y)| + \binom{s}{s/2}$.
  Thus, for all $y \in \integer_K$, we have that
  $|f^{-1}(y)|$ is within $\binom{s}{s/2}$ of the
  average value of $\frac{2^s}{K}$. Hence, 
  \begin{equation*}
    \binom{s}{\frac{s}{2}} \leq \frac{4}{5} \frac{2^s}{K} \frac{K}{\sqrt{s}}
    \le \frac{4}{5} \left( \left| f^{-1}(y) \right| 
      + \binom{s}{\frac{s}{2}} \right) \frac{K}{\sqrt{s}}
    \enspace ,
  \end{equation*}
  from which follows 
  \begin{equation*} 
    \binom{s}{\frac{s}{2}} \le \frac{4}{ 5 \frac{\sqrt{s}}{K} - 4 }
    \left| f^{-1}(y) \right|
    \enspace .
  \end{equation*}
\end{proof}
\begin{lemma}\label{lm:identicalsize2subsets}
  Let $B_1 = \cdots = B_s = \{0,b\}$ be $s$ identical size-$2$ subsets
  of $\integer_T$, with $s \geq T^2$.  Then the multiset $B_1 + B_2 +
  \cdots + B_s$ has bias at most $4 |H|/{s^{1/2}}$ with
  respect to the subgroup $H=\langle b \rangle$.
\end{lemma}
\begin{proof}
  Set $K = |H|$ and define function $f: \{0,1\}^s \rightarrow
  \integer_K$ by $f(a_1, \ldots, a_s) = \big(\sum_{i=1}^s a_i\big)
  \textup{ mod $K$}$.  Then we may generate the multiset $B_1 + B_2 +
  \cdots + B_s$ as $b \cdot f(\{0,1\}^s)$.  Applying
  Lemma~\ref{lm:coins} gives that $f$ is almost unbiased on
  $\integer_K$ and hence $b\cdot f$ is almost unbiased with respect
  to~$H$.
\end{proof}
\begin{lemma}\label{lm:size2subsets}
  Let $B_1, \ldots, B_r$ be size-$2$ subsets of $\integer_T$, with $r
  \geq T^3$.  There exists a nontrivial subgroup $H \subgroup
  \integer_T$ such that $B_1 + B_2 + \cdots + B_r$ has bias at most $4
  T^{3/2}/ r^{1/2}$ with respect to~$H$.
\end{lemma}
\begin{proof}
  First suppose $0 \in B_i$ for all~$i$.  There exists some nontrivial
  element $b \in \integer_T$ such that $B_i = \{0,b\}$ for $s$ of the
  subsets, with $s \geq r/T \geq T^2$.  Applying
  Lemma~\ref{lm:identicalsize2subsets} on these $s$ subsets yields a
  multiset of bias at most $4 |\langle b\rangle|/s^{1/2} \le 4
  T^{3/2}/r^{1/2}$ with respect to $\langle b \rangle$.  By
  Lemma~\ref{fact:adding}, adding the remaining $r-s$ subsets to this
  multiset does not increase the bias.
  
  In general, we do not have that $0 \in B_i$ for all~$i$. In this
  case, observe that by Lemma~\ref{fact:adding}, adding any offset to
  a multiset does not change its bias, and thus we may reduce to the
  former case by adding an appropriate offset $d_i$ to subset $B_i$
  such that $0 \in B_i + \{d_i\}$, for each~$i$.
\end{proof}
\begin{proof}[Proof of Theorem~\ref{thm:addition}]
  Let $B_i \subseteq_R A_i$ be a random size-2 subset of $A_i$, for
  each~$i$.  By Lemma~\ref{lm:size2subsets}, the sub-rectangle $R' =
  B_1 \times \cdots \times B_r$ is almost unbiased with respect to
  some nontrivial subgroup~$H'$.  Since $H'$ is nontrivial, it
  contains $H=\{0,2^{t-1}\}$, and hence $R'$ is also almost
  unbiased with respect to~$H$. By this selection process, every $(a_1,
  \ldots, a_r) \in A_1 \times \cdots \times A_r$ has the same
  probability of being selected and, hence, $R$ itself is almost
  unbiased with respect to $H$.
\end{proof}
\begin{proof}[Proof of Lemma~\ref{lm:unbiased}]
  Set $t = \frac{1}{6} \log n$ and $T = 2^{t}$.  Consider any
  rectangle $R=A_1 \times \cdots \times A_n$ involving at least $r
  \geq n^{5/6} = T^5$ parties.  By~the Addition Theorem, the multiset
  $A_1 + \cdots + A_n$ has bias at most $\bigO(T^{3/2}/r^{1/2}) \subseteq
  \bigO(1/n^{1/6})$ with respect to $\{0,2^{t-1}\}$.  Hence, rectangle
  $R$ has bias at most $\bigO({1}/{n^{1/6}})$, too.
\end{proof}

\section{Conclusions}\label{sect:concl}\label{sect:nonlocconcl}

We studied experiments for validating quantum
nonlocality in the presence of noise and with imperfect detectors.
Specifically we concentrated on the generalization of the GHZ paradox
to $n$ parties previously considered as a quantum communication
complexity problem \cite{bdht:multiparty}.  

There are several directions in which one may wish to improve the
result Eq.~(\ref{REL}). The first concerns the evaluation of the
right-hand side of this relation.  A detailed investigation of the
proof shows that the right-hand side becomes nontrivial only for
values of $n$ that exceed a few hundred.  Therefore our result will
not be useful for the moderate values of $n$, say, $n \leq 10$, which
may be attainable by real-world experiments in the next few years. It
would be interesting to try to improve Eq.~(\ref{REL}) so as to make
it relevant for small values of $n$. 
Can the gap between the result in the absence of noise (when the
right-hand side is $\bigO(n^{-1})$) and the result in the presence of
noise be closed?

Another question concerns our notion of error, which is not entirely
appropriate to a multiparty setting: one expects that
each party may induce an error independently of the other parties.
Thus it would be more natural to consider that the probability of an
error goes as $\epsilon = 1 - \delta^n$. We do not know whether a
constraint of the form Eq.~(\ref{REL}) holds in this case.

Notwithstanding the above directions in which improvements are
possible, there is a specific sense in which the above result can be
shown to be close to optimal. Consider $n$ parties
who share an entangled state $\ket\psi$ of dimension $2^n$. 
Each party's system is two
dimensional, \ie each party has a single qubit. 
Fix a total-variation distance $\epsilon_\textup{var}$. Then
for any measurement scenario involving local measurements on the
quantum state $\ket\psi$, the amount of (super-luminal) communication
required to reproduce these correlations up to total-variation distance $\epsilon_\textup{var}$ is at most of order $n \log
n$, and the maximal detector efficiency $\eta^*$ for which these
correlations are local is of order $n^{-c}$ for some constant
$c$. This result will be reported elsewhere~\cite{massar04:simulatingNonlocality}. 
It shows that the example
considered above is close to
maximally nonlocal, at least if one restricts oneself to a large
number of parties each possessing a single qubit.

\bibliography{preamble,body}

\end{document}